\documentstyle[12pt]{article}
\newcommand{\be}{\begin{equation}}
\newcommand{\ee}{\end{equation}}
\newcommand{\ba}{\begin{array}}
\newcommand{\ea}{\end{array}}
\newcommand{\beqa}{\begin{eqnarray}}
\newcommand{\eeqa}{\end{eqnarray}}

\renewcommand{\ln}{{\rm ln}}

\newcommand{\matr}{\left( \begin{array}}
\newcommand{\ematr}{\end{array} \right)}

\newcommand{\smm}{{singlet Majoron model}} 
\newcommand{\SM}{{Standard Model}}
\newcommand{\lsim}{\;\raise0.3ex\hbox{$<$\kern-0.75em\raise-1.1ex\hbox{$\sim$}}\;}
\newcommand{\gsim}{\;\raise0.3ex\hbox{$>$\kern-0.75em\raise-1.1ex\hbox{$\sim$}}\;}
\def\be{\begin{equation}}
\def\ee{\end{equation}} 
 
\begin{document}

\begin{titlepage}
 
\mbox{}\hfill\makebox[4cm][l]{TURKU-FL-P10}\newline
\mbox{}\hfill\makebox[4cm][l]{(hep-ph/9404268)}
\vfill
 
\Large
 
\begin{center}
{\bf  Vacuum stability in the \smm }
 
\bigskip
\normalsize
{ J.\ Sirkka and I.\ Vilja}\\[15pt]
{{\it Department of  Physics, University of Turku, 20500 Turku, Finland}}
 
{April 1994}
%\maketitle
 
\bigskip
 
%\today
 
\vfill
 
\normalsize
 
{\bf\normalsize \bf Abstract}
 
 \end{center}
 
\normalsize
 
We study the vacuum stability of the singlet Majoron model using full 
renormalization group improved scalar potential and Monte Carlo techniques.
We show that in the perturbative regime of the various free parameters, 
the vacuum stability requirement together with LEP limits is passed by
18\% of the parameter 
space if the scale of new physics is 10 TeV and  6\% if the scale is $10^{14}$ 
GeV. Moreover, if the baryogenesis condition for scalar couplings is required, no
portion of the parameter space survives.   
 
\end{titlepage}
   
\newpage

\setcounter{page}{2} \noindent%{\it 1. Introduction. } 

The \smm\ \cite{CMP} as one of the simplest extensions of the Standard Model
(SM) offers an explanation to many problems that remain open in the minimal
\SM. In addition to the usual doublet scalar field of the \SM\  
the \smm\ contains a complex, electroweak singlet scalar
field. Also, right--handed electroweak singlet neutrinos are introduced
to the model so, that an extra, global $U(1)$ symmetry appears.
This global symmetry is broken approximately at the electroweak scale
manifesting that the other of the new scalar degrees of freedom
 becomes massive whereas the other remains as a massless Goldstone boson. 
The non-zero vacuum expectation value of the singlet field then implies
a mass term for the right--handed neutrinos. 

A strong pro in favour to such an extended model of the \SM\ is that the
weak scale baryogenesis, which appears to be somewhat problematic in the \SM\
\cite{KRS,Shapo}, is easier to realize in the \smm\ \cite{EKV,Vilja}.
A problem with the \SM\ is that the experimental lower bound for the
Higgs boson mass \cite{Mori}, $m_H > 60$ GeV, is too high compared to its
 theoretical 
upper bound in order to avoid the erasure of the baryon number by sphaleron
\cite{M&KM} mediated transitions \cite{BKS}. In the Majoron model these
problems are circumvented by introducing the new scalar degrees 
of freedom \cite{EKV,KUY}. Moreover, the \smm\ together with the
sew--saw mechanism  offers an explanation to the vanishingly
small  (left--handed) neutrino masses. By--products of the sew--saw
mechanism, the heavy  right--handed neutrinos may also serve as a source
for cosmological baryon asymmetry due  to their lepton number violating
decays which can be converted  to baryon number by  sphalerons
\cite{Vilja}. For a review of some other baryon number production mechanisms 
applicable in the \smm, see \cite{CKN}.

 To be a realistic model of electroweak 
scale physics, any model has to have the vacuum of the observed universe
 stable enough. If the stability is not achieved,
the theory can not be fully correct. This stability is, however needed only up 
to some scale, like supersymmetry scale or GUT scale, where new interactions becomes 
important: new phenomena appears and saves the stability. In this letter we study 
the stability 
properties of the \smm\ at zero temperature. We work using  one--loop
perturbation theory and make a full  renormalization group (RG) improved
stability analysis of the scalar potential. Such an analysis for the \SM\ 
has first been performed by Flores and Sher \cite{FS} and recently by
Sher  \cite{Sher} for new. They found that the stability of the
potential requires that  the Higgs mass $m_H > 75\ \mbox{{\rm GeV}} +
1.64(m_{top} - 140\ \mbox{{\rm GeV}})$. For top quark  mass 174
GeV, which we use as an example, $m_H > 131$ GeV. Renormalization group 
improved stability analysis
for two  doublet model has also been carried out \cite{Komatsu&FS2}. The \smm\ 
case differs from both above  mentioned cases, not only because of
different scalar content, but also because of the inclusion of the
right--handed neutrinos. Their 
unstabilizing effect to the potential may be remarkable like the effect of 
the top quark has in the minimal \SM. 

The tree--level potential of the \smm\  reads 
\be V_0(H,S) = m_H^2 |H|^2 +
m_S^2 |S|^2 + \gamma |H|^2|S|^2 + \beta |S|^4 + \lambda |H|^4, 
\ee
where, in a spontaneously broken theory, the mass--like parameters
$m_H^2$ and $m_S^2$ are negative. The couplings $\beta$ and $\lambda$
are positive with $\gamma^2 < 4\lambda \beta$ which guarantees that the
tree--level potential has a stable, non--trivial  minimum. (See
Ref. \cite{EKV} for a detailed study of the potential.) The right--handed 
neutrinos couple to the singlet field with Yukawa couplings $g_i$
($i = 1,\, 2,\, 3$)  corresponding to the three generations of leptons. 
In practise we,
however, consider only the heaviest  right--handed neutrino and
disregard the other two. The moduli of vacuum expectation values  $\sqrt
2 f$ and $\sqrt 2 \bar f $ of the scalar fields $H$ and $S$,
respectively, are given by 
\begin{eqnarray} f^2 &= &{-2\gamma m_S^2 +
4\beta m_H^2\over \gamma^2 - 4\lambda\beta}, \\ \bar f^2 &= &{-2\gamma
m_S^2 + 4\lambda m_H^2\over \gamma^2 - 4\lambda\beta}, 
\end{eqnarray}
and the mass eigenstates are 
\be m_\pm^2 = \lambda f^2 + \beta\bar f^2 \pm
\sqrt D, 
\ee where $D = (\lambda f^2 - \beta\bar f^2)^2 +
\gamma^2f^2\bar f^2$.

The full set of the renormalization group equations needed to stability analysis
calculated using one--loop perturbation theory reads 
\begin{eqnarray} 
\dot \alpha_s &= &-{7\over 2\pi}\alpha_s^2,\\ 
\dot \alpha   &= &-{19\over 48\pi}\alpha ^2,\\
\dot \alpha'  &= &{41\over 48\pi}\alpha'^2,\\ 
\dot \alpha_t &= &{1\over 2\pi}
[\frac 92 \alpha_t^2 - 8\alpha_s\alpha_t - \frac 94 \alpha\alpha_t
- \frac {17}{12}\alpha'\alpha_t],\\ 
\dot \alpha_i &= &{9\over 4\pi} \alpha_i^2;\qquad i=1,2,3,\\ 
\dot \lambda  &= &4\lambda\gamma_H
+ {1\over 8\pi^2}[B + 12\lambda^2 + \frac 12 \gamma^2],\\ 
\dot \beta &= &
4\beta\gamma_S + {1\over 8\pi^2}[B' + 10\beta^2 + \gamma^2],\\ 
\dot\gamma &= &
2\gamma[\gamma_H + \gamma_S] + {1\over 4\pi^2}[\gamma^2 +
2\gamma\beta + \frac 92 \gamma\lambda ], 
\end{eqnarray} 
where $\alpha_A=g_A^2/(4\pi)$ for the gauge as well as for the Yukawa
couplings. The parameters $B$ and $B'$ are functions of couplings
defined by
\begin{eqnarray}  
B &= &3\pi^2 \left [{3\alpha^2 +
2\alpha\alpha' + (\alpha')^2\over 4} - 4\alpha_t^2\right ],\\ 
B'&= &- 6\pi^2\left [ \alpha_1^2+\alpha_2^2+\alpha_3^2\right ]. 
\end{eqnarray} 
The anomalous dimensions of the scalars 
\begin{eqnarray} 
\gamma_H &=&
{3\over 16\pi} [-3\alpha - \alpha' + 4\alpha_t],\\ \gamma_S &= &{3\over
4\pi} [\alpha_1+\alpha_2+\alpha_3]  
\end{eqnarray} 
are also needed. The dots stand for
the derivative with respect to the logarithmic scale variable $t = \ln 
(\mu / M_Z)$,  where $\mu$ is the 
effective scale. %identified with $[(|H|^2 + |S|^2)/2]^{1/2}$.
We renormalize the couplings as $\alpha_s(0) = 0.114  ,\
\alpha(0) = 0.0335  ,\ \alpha'(0) = 0.0102  ,
\  \alpha_t(\mu = 2m_{top}) = m_{top}^2/(2\pi f^2) = 0.0790$ 
for $m_{top} = 174$ GeV. The right--handed neutrino couplings are 
renormalized using \cite{GP} 
\be
\alpha_i(\mu = 2 m_{N_i}) = {1\over 2\pi}{m_{N_i}^2\over \bar f^2}, 
\ee
where $M_{n_i}$ are the physical heavy neutrino masses. The vacuum stability 
requires now \cite{Sher2} that inequalities $\beta (t) > 0$, $\lambda (t) > 0$
 and $\gamma
 (t)^2 < 4\beta (t)\lambda (t)$ holds for all $t$ up to the given scale of new 
physics. Note, that the quadratic terms of the scalar potential are not important
because we are only interested in large scales.

The large number of degrees of free parameters makes the integration of 
Eqs.\ (5) - (12) complicated and difficult to visualize in practise.
Therefore, to study the RG--equations, we perform a Monte Carlo analysis 
where the initial values  of the parameters 
$\lambda$, $\beta$, $\gamma$, $\bar f$  and the largest 
neutrino
Yukawa coupling $g_Y$ are generated randomly at $\mu = M_Z$. The logarithms
of these parameters are taken to be uniformly distributed, so that 
we are able to cover
the whole parameter space rather densely. As mentioned, we neglect the other
right--handed neutrino Yukawa couplings but the largest one. This does not
invalidate the analysis due to the structure the Yukawa couplings emerge in the
RG--equations. The ranges of the parameters are 
\be 
10^{-3}<\beta,\gamma<10^{-1},\qquad 5\times 10^{-3}<\lambda<10^{-1}
.\ee
These bounds are due to the requirements that the baryogenesis analysis
of \cite{EKV} and the perturbation theory are applicable. Moreover we choose
\be 
1 \ \mbox{\rm GeV}<\bar f<10 \ \mbox{\rm TeV},\qquad 10^{-3}<g_Y<1,
\ee
where the lower bound for $\bar f$ comes from the experimental upper bound and  
the see--saw mechanism estimate for
the mass of the electron neutrino $m_{\nu_e}\sim m_e^2/(g_Y\bar f)$. 
On the other hand $\bar f$ cannot be much larger than $f=246$ GeV because 
the singlet scalar would effectively  decouple \cite{EKV}. (This kind of 
model is, of course, technically allowed, but is physically not very 
meaningful containing a new, {\em ad hoc} symmetry breaking scale.) 
The range of
$g_Y$ was taken to be on the perturbative domain but not too small for
the second order corrections in $\beta$, $\lambda$ or $\gamma$ to be
effective.  

From the randomly chosen initial values only those  are accepted 
for which the two conditions,
$4\beta\lambda > \gamma^2$ at $\mu=M_Z$ and there is no Landau
singularity below the maximum scale studied, holds. Approximately $73
\% $ of the generated values pass this test when maximum scale is the
supersymmetry scale $\mu=10$ TeV and $71 \%$ when the maximum scale
is the unification scale $\mu=10^{14}$ GeV.
Furthermore three different cuts are applied to limit the parameter space:
\begin{enumerate}
\item It is required that vacuum is (absolutely) stable up to the given
maximum scale. In the case of supersymmetry scale 24 \% of the generated
points in the parameter space pass this cut and in the case of
unification scale 8 \% pass the cut. 
\item It is required that the scalar masses calculated from the
generated point are compatible with the model independent LEP lower 
limit for a scalar particle mass. This means (see \cite{EKV} for
details) that 
\be
\cos^2z\,\Gamma(m_+)BR_++\sin^2z\,\Gamma(m_-)BR_-<\Gamma(60 \ \mbox{\rm
GeV}), \label{lep1}
\ee
where $\Gamma(m)$ is the decay rate of $Z$ to a fermion pair and a scalar
with mass $m$, $BR_\pm$ are the branching ratios of scalars to ordinary
fermions and $z$ is the angle of rotation from current to mass
eigenstates. 30 \% of the points pass this cut. 
\item  Supplementary to the second limit there is another limit from LEP
data. The latter relies on a event topology analysis of the reaction
$Z\rightarrow Z^*H_0\rightarrow l^+l^-(\nu\bar\nu)$ when the former
relies on observing reduced number of the standard fermionic decays of
this reaction \cite{EKV}. The resulting limit reads
\be
\cos^2z\,\Gamma(m_+)+\sin^2z\,\Gamma(m_-)<4\times\Gamma(60\ \mbox{\rm
GeV}).\label{lep2}
\ee
Also this cut is passed by 30 \% of the points.
\end{enumerate}
 
In Figure 1 the resulting projection of the parameter space in
the $m_+$--$m_-$ --plane is shown after applying three combinations
of the above cuts. Note, that these masses are not really the physical ones, 
because their scale dependence is not taken into account. As the scalar masses are not
too far from the weak scale $M_Z$ the correction
due to the running of the masses may, however, be neglected. 
The parameter space is reduced to 18 \% after
applying all cuts in the case of the supersymmetry scale and to 6 \% 
in the case of the unification scale. The vacuum stability bounds $\lambda$ to be 
larger than 0.034 (0.067) for $\mu = 10$ TeV ($10^{14}$ GeV) except some very few 
points which cover less than 0.03\% of the parameter space. Therefore it can be 
concluded that vacuum
stability requirement and baryogenesis bound $\lambda\lsim 0.018$ \cite{EKV} are
somewhat controversial. Applying the vacuum stability cut and the baryogenesis
bound for $\lambda$ together leaves zero parameter space. It is also
noteworth that Monte Carlo analysis shows that most of the values of the
neutrino masses passed by  the vacuum stability and LEP limits are
between $20$ GeV and some hundred GeV's, i.e. not very small nor very
large neutrino masses are favoured.

Our results infer that the vacuum stability and baryogenesis bounds in the \smm\  
can not be fitted together. However, both limits have some uncertainty. Because the
baryogenesis bound may be affected by non--perturbative effects similar to ones 
in the \SM, the upper bound for the doublet self--coupling $\lambda$ may be relaxed.
Yet, if our vacuum is not absolutely stabile but its
life--time exceeds  the age of the present universe $\approx 10^{10}$ yr, the vacuum
stability bound will be somewhat relaxed, too. (For a general discussion, see 
Ref.\ \cite{Sher2} and references therein.) In the \SM\  the possibility of an
un--absolute vacuum allows, however, only about 10 GeV smaller higgs mass than the
absolute stability requirement \cite{Sher}. It is not probable that this effect 
would be large either in the \smm. Taking into account 
the uncertainties of the bounds it might be possible that a very thin right--angular
zone beginning at $(m_-,m_+)\approx (0,90)$ GeV and ending at $(m_-,m_+)
\approx (90,150)$ GeV survives both baryogenesis and vacuum stability bounds, i.e.\
the gap between $\lambda < 0.018$ and $\lambda > 0.034$ can be filled.

%\newpage

%\newpage
\noindent {\bf FIGURE CAPTION}

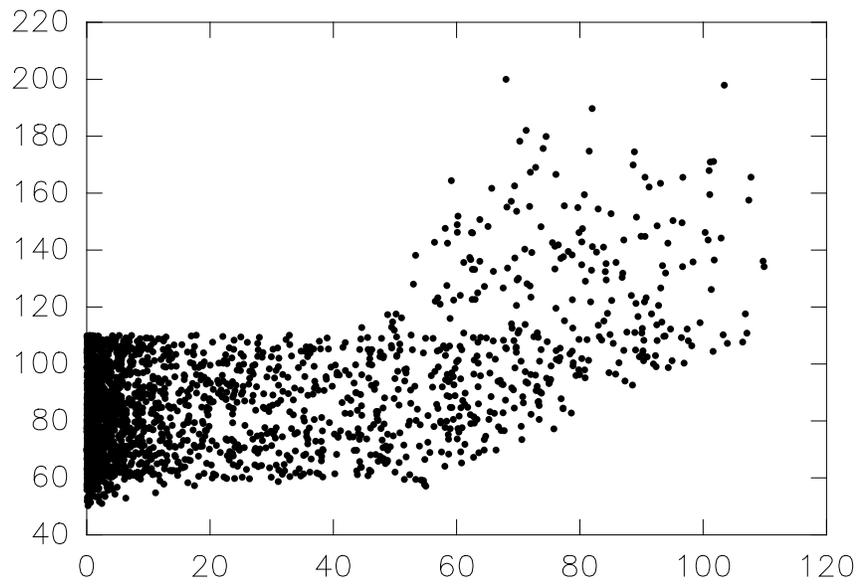
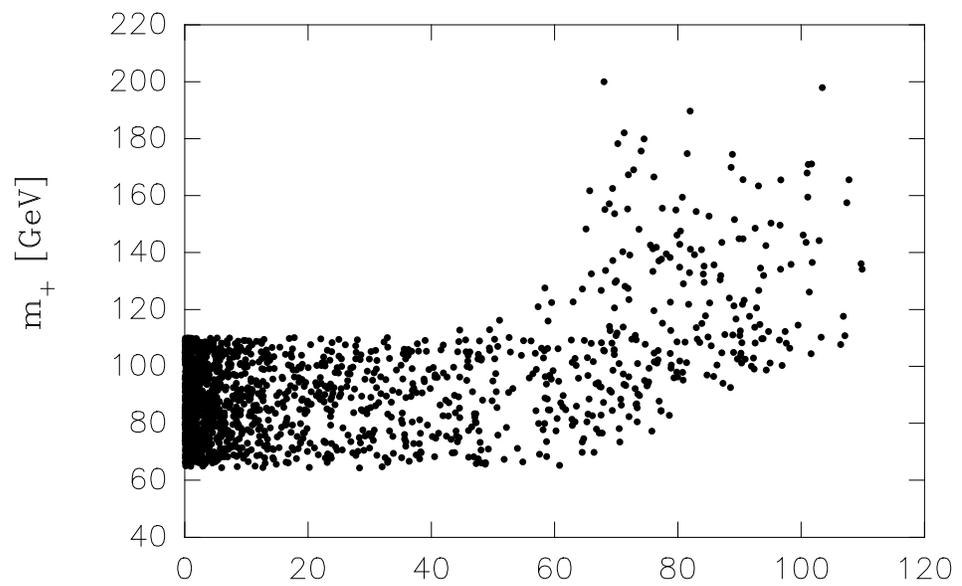
\noindent {\bf Figure 1.} Distribution of randomly generated points in
the $m_- - m_+$ --plane after applying combinations of cuts. {\bf a)}
Points (26 \% ) which fulfill LEP limits (\ref{lep1}) and (\ref{lep2}). {\bf b)}
Points which pass all cuts when the scale of the new physics is $\mu=10$
TeV. {\bf c)} Same as b) for $\mu=10^{14}$ GeV. 

%\noindent {\bf Figure 2.}

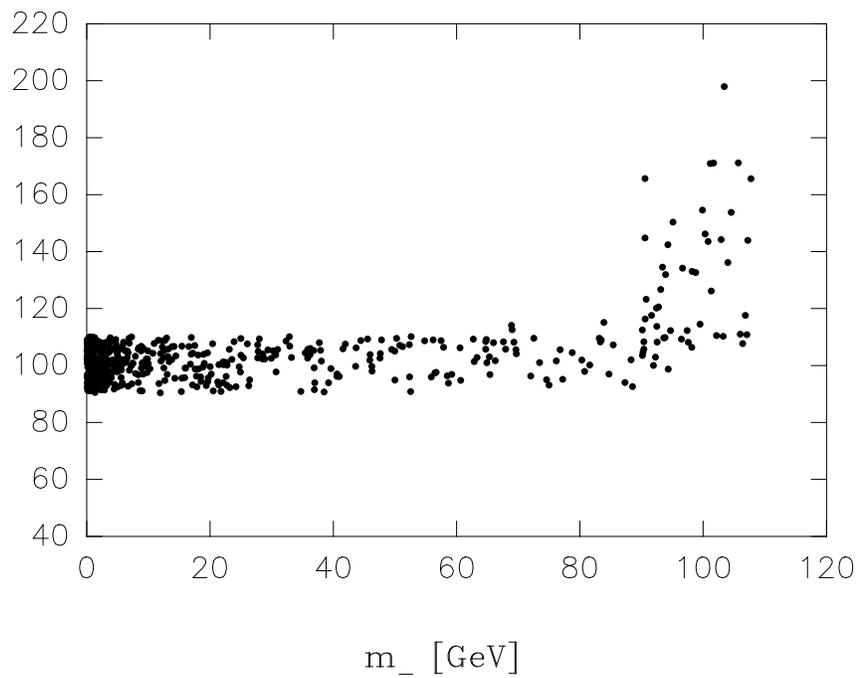
\end{document}